# Progress in the Development of a Plasma Panel Detector

R. Ball[1], J.R. Beene[2], Y. Benhammou[3], M. Ben Moshe[3], J. W. Chapman[1], T. Dai[1], E. Etzion[3], P.S. Friedman[4], D.S. Levin[1], Y. Silver[3], G. Sherman[3], R.L. Varner Jr.[2], C. Weaverdyck[1], S.White[5], J. Yu[1], B. Zhou[1]

*Abstract–* **Plasma Display Panels (PDP), the underlying engine of panel plasma television displays, are being investigated for their utility as radiation detectors called Plasma Panel Sensors (PPS). The PPS a novel variant of a micropattern radiation detector, is intended to be a fast, high resolution detector comprised of an array of plasma discharge cells operating in a hermetically sealed gas mixture. We report on the PPS development effort, including recent laboratory measurements.**

## INTRODUCTION

We are investigating an ionizing particle detector technology based on Plasma Display Panels (PDP). The PDP's are the principal component of flat panel plasma television displays. Their design and production is supported by an industrial infrastructure with four decades of development. As televisions and displays, plasma panels have proven reliability, durability and very long lifetimes, coupled with low costs with applications in both the commercial and military sectors. Our objective is to develop from the PDP technology, a novel micropattern radiation detector – herein referred to as Plasma Panel Sensors (PPS), the fundamental concept of which was initially introduced earlier (references 1-5). The PPS is intended to benefit from many of the key attributes of plasma panels.

This technology offers the potential of high gain, fast response times, a commensurate high data rate capability and low power consumption. The RMS spatial resolution will be set, in part, by the dimensions, pitch and the uniformity of electrodes on the display glass substrate. Current low cost manufacturing capability allow for electrodes with as little as 1-2 micron non-uniformity, well below the order of 100 micron pixel sizes intended for PPS detectors. Here we provide an overview of the PPS concept and report on initial laboratory test results.

I. PLASMA PANEL SENSORS DETECTOR OVERVIEW

The PPS is based on the PDP, which comprises millions of cells per square meter, each of which can, when provided with a signal pulse, initiate and sustain a plasma discharge. As a plasma panel detector, a PDP cell can be biased to discharge when a free-electron is generated or injected into the gas. Therefore, the PPS is intended to function as a 2D array of pixel-sensor-elements or cells, each independently capable of detecting single free-electrons generated within the cell by incident ionizing radiation. The basic element of PDPs consists of either orthogonal (i.e. columnar-discharge) or collinear (i.e. surface-discharge) arrays of electrodes deposited onto glass substrates, separated by a gas discharge gap. PDPs have been manufactured as DC and AC units. Televisions employ the latter while some limited use displays, as will PPS units, use a DC mode. The PDP cells are on the order of 200 μm in each dimension for HDTV, although cell dimensions on the order of 100 μm have been made for military applications. Because of the small electrode gaps, large electric fields arise with only a few hundred volts of bias. The plasma discharges are usually made in Penning mixtures of noble gases: typically Xe and Ne gas at about 500 torr. It is important to note that the gas in a PDP is permanently and hermetically sealed in the panel's glass envelope. Signals reported in this note include those obtained on a display panel that has been sealed for over seven years.

The plasma panel detector is biased to discharge when a radiation induced free-electron is generated in the gas. Such electrons then undergo rapid electron multiplication resulting in an avalanche and discharge that can be confined to the local pixel cell space. As in PDP products, this process is self-limiting and self-contained by various means, one of them being a localized impedance at each cell. The total charge available to produce a signal is that stored by the cell's effective capacitance and limits the maximum gain. Since the cell is operated above the proportional mode it may be considered to be a micro-Geiger counter. The signal pulse will be independent of the number of initiating free electrons, rendering the PPS as intrinsically digital. Depending on the pixel geometry, panel design and gas, the gain can be sufficiently large to produce signals with volt scale amplitudes. Discharge confinement and localization is achieved by various methods, such as a pattern of dielectric material deposited over the bare metal electrodes to establish regions were discharges can occur.

1. University of Michigan, Department of Physics, Ann Arbor, MI
2. Oak Ridge National Laboratory, Oak Ridge, TN
3. Tel Aviv University, Beverly and Raymond Sackler School of Physics and Astronomy, Tel Aviv, Israel
4. Integrated Sensors, LLC, Ottawa Hills, OH
5. Brookhaven National Laboratory, Upton, NY

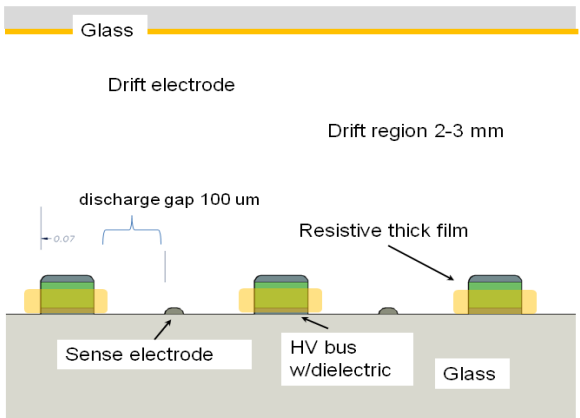

**Figure 1:** 2D view of conceptual representation for test device substrate. Pixels formed by gap between HV (discharge) and sense lines. Quench resistances formed by resistive deposition. Signals form on sense electrodes.

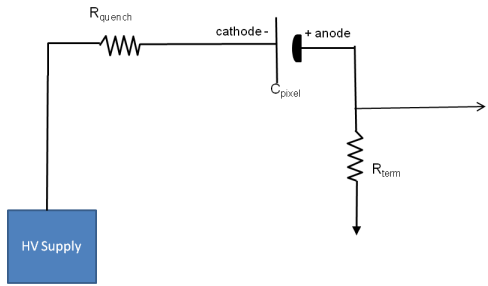

**Figure 2**: Simplified schematic of single PPS pixel. The quench resistance terminates the discharge and is chosen sufficiently large (0.1 to several mega-ohms) to prevent sustained discharges.

A. Cell Geometry

The cell geometry includes the dimensions of the electrodes, their pitch and vertical spacing, and dimensions of cell gas drift gap. A representation of the PPS electrode layout is shown in Figure 1. For minimum ionizing particle (MIP) detection the electrode layout should have a "vertical" drift region on the order of 2 to 3 mm and a transverse electric field avalanche region of ~100 μm. The drift region is required to ensure sufficient probability that a passing particle will produce at least one ion-pair. Due to the large gain, a single ion-pair can initiate a signal, suggesting that a drift region of ~ 3 mm is sufficient to produce signals with very high efficiency (5).

B. Electrical Characteristics

In addition to Figure 1 a conceptual representation of a cell configuration is shown in reference (5). The cell is defined by a local electrode arrangement with an intrinsic capacitance and an embedded resistance in the high voltage feeds. The discharge gap is determined by the proximity of the two electrodes, the high voltage (or discharge) side of each being fed by a resistance. This local impedance effectively terminates and helps to localize the discharge. The effectiveness of this resistance to electrically isolate the discharge is investigated with SPICE (6) simulations. A simplified schematic of a single pixel is shown in Figure 2 and a chain of pixels, including line resistances and stray capacitances is shown in Figure 3. Represented in this schematic are the embedded pixel resistances, the pixel capacitances, stray capacitances and the termination resistance. The various capacitive couplings are modeled with COMSOL (7). Also included is the discharge current source - which is assumed to occur on a ns timescale. The simulations in reference (5) indicated that the neighboring cells to the discharging cell experience no change in their bias; they remain charged independently ready to respond to incident radiation. The voltage transients on neighboring pixels will be measured in future lab tests.

In the projected side view of Figure 1, the resistors are directly embedded in a thick-film process, and are represented by the light green regions separating the high voltage bus from the discharge electrodes. The resistors do not occupy sensitive detection area on the substrate. Also shown (not to scale) is the 2-3 mm drift field. The electric field, where wires represent the electrodes, is shown in Figure 4, computed with Garfield (8) and also with COMSOL (7). While the dimensions shown are illustrative only, the 100 μm gaps shown are easily within current low-cost manufacturing capability. The depth of the electrode into the page sets the resolution of an orthogonal coordinate and also affects the total pixel capacitance. The extent of this depth is being investigated. It depends, in part, on the resolution requirements of a 2$^{nd}$ coordinate.

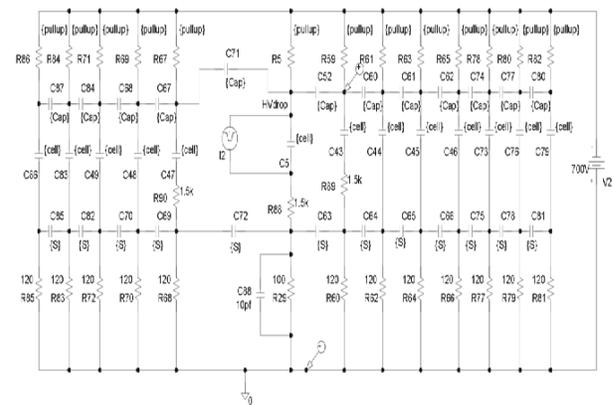

**Figure 3:** Schematic view of a pixel chain with discharge regions represented by capacitors isolated and quenched by embedded resistors. The modeling includes stray capacitances that couple discharge electrodes to nearby neighbors

II. LABORATORY TESTS OF PDP PANELS

An experimental program has been initiated to establish the overall feasibility to induce self-limiting, plasma discharges and to extract initial hit rate measurements for well-defined geometries, gas mixtures and electric field configurations. These first signal measurements were conducted using sealed, thin gap envelope glass PDP panels, fabricated and gas filled in August 2003 (1)(9) with a Penning mixture of 99% Xe, 1% $O_2$ to 650 torr. These modules consists of two float glass 6 cm x 13 cm plates active area (1/8 inch thick) each with thin-film strips of metallic and transparent electrodes with approximately 1 mm pitch and oriented orthogonally. The plates are separated by a gas gap of approximately 220 microns. Signals were induced using a $^{90}$Sr beta source. Additional new panels, having an open gas port, were also tested. The gas port enables trials with different gas mixtures.

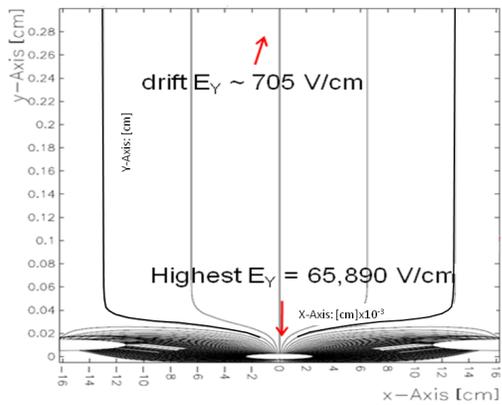

**Figure 4:** 2D projection of electric field drift lines for a geometry of the type similar to Figure 3. (Note: different scales on X and Y axes.)

Figure 5 shows the basic configuration used to obtain the signals reported here. The intersection of the electrodes comprises a columnar discharge region across the gas gap. The quench resistance, depending on the panel used, ranges from $10^5$ ohms to several mega-ohms. This resistance is selected sufficiently high to ensure that sustained discharge does not occur. Termination is approximately 50 ohms. Pulses are induced by bringing a collimated 3.7 mCi $^{90}$Sr source to the panel pixel. The computed spectrum using Geant4 is shown for those βs that have passed through the front cover glass and enter the gas. Calibration of the actual number of βs entering a given pixel region is not yet complete, but the number is low and therefore the β source acts like a single particle generator with rate of order Hz. The signal is directed to an oscilloscope, discriminator and scalar.

For the signal reported in Figure 6, using the small gap Xe filled panel, the bias voltage was ramped up to near the predicted breakdown voltage (600 V) based on the Paschen parameter for pure Xe (10) (about 13 torr-cm), then incremented slowly until pulses were observed. (The presence of $O_2$ in the mixture is presumed to be non-existent as it is readily absorbed in the glass and by the thick-film Ni electrodes). The main features of this signal are its volt scale amplitude and fast rise time (10%-90%) ~ 2 ns. The overshoot and ringing are presumed to result from residual impedance mismatches.

Initial tests using a newly fabricated PDP panel with an accessible gas port, a 340 μm gas gap and 1.5 mm electrodes were conducted. This panel was evacuated and filled to 700 torr with an Argon based mixture.. Like the Xe filled unit, this panel proved to be sensitive to the $^{90}$Sr source yielding volt scale signals into a 50 ohm termination. We measured single pixel hit rates vs. high voltage with and without the pixel exposed to the collimated source. We also investigated discharge spreading wherein a hit in one pixel could produce a spurious hit in a neighbor. Figure 7 shows a measurement of the single pixel hit rate as a function of the high voltage, with and without the source applied to the pixel. The turn on of the pixel occurs at ~840 V. Both signal (red) and background (blue) rates are observed to increase with HV although the signal rate remains always above background. At higher voltages (e.g. above 960 V) the pixel tends to exhibit a sustained discharges or multiple hit bursts.

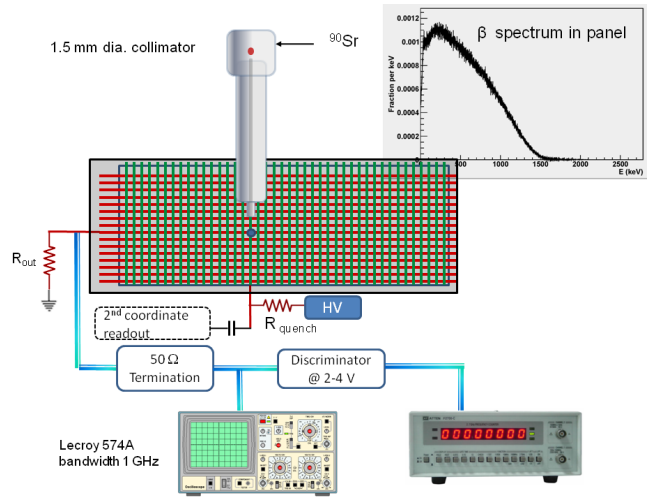

**Figure 5:** Layout of laboratory demonstration of principle tests. The $^{90}$Sr collimated source injects electrons into a pixel defined by the intersection of electrodes.

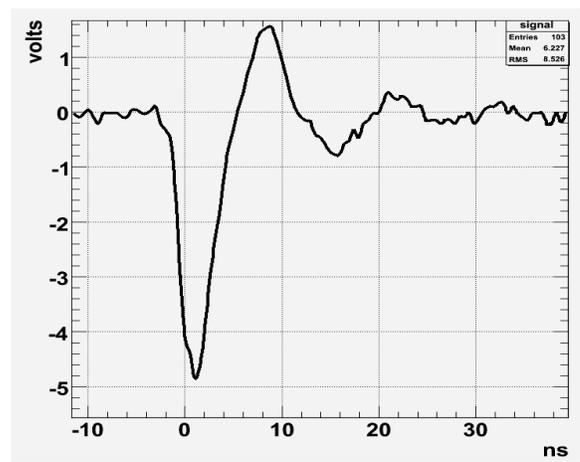

**Figure 6:** Example of unamplified signal pulse from a Xe filled panel. The rise time (10%-90%) is ~ 2 ns, and reflects the combined time constants of the cell, stray capacitances and scope bandwidth.

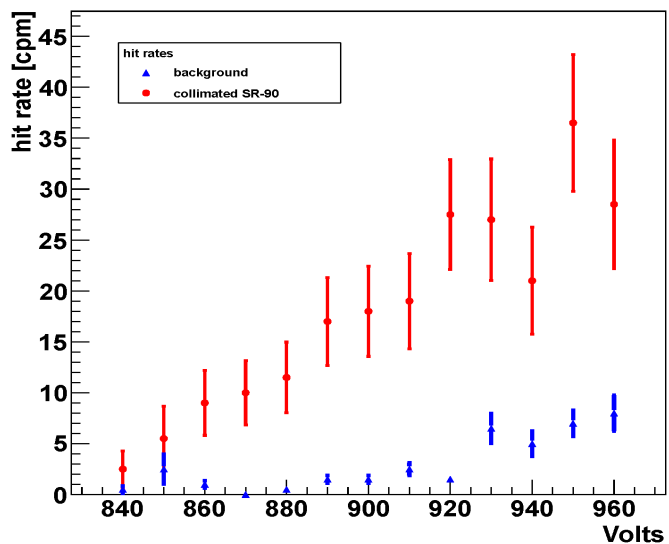

**Figure 7:** Single pixel count rates with/without a $^{90}$Sr β source applied to a single pixel of a PDP panel filled with a 99% Ar gas mixture at 700 torr.

Discharge spreading from a hit pixel to its nearest neighbors could be a potential source of spurious hits in an open structured PPS detector configuration. To investigate this in the PDP panels, the signal lines from several pixels on each side of a target pixel were monitored on an oscilloscope that was triggered by a target pixel. The neighbors did exhibit small positive pulses whose amplitudes were about 20% of the target pixel negative signal pulse. These pulses are expected however, due to a combination of inductive and capacitive coupling of the hit lines to its neighbors and do not constitute discharge spreading. With several thousand individual signal pulses on a target pixel, the fraction of those producing spurious pulses on the four nearest neighbor pixels was approximately 4%. This still very preliminary result does suggest that for at least the first panel tested, discharge spreading does not appear to be very significant.

### III.  SUMMARY

A program to develop plasma panel radiation detectors using plasma television display technology is underway. We have instrumented single pixels from small display units, filled with gas seven years earlier, and from newly fabricated devices. Fast signals, externally induced by a $^{90}$Sr source have been observed in both panels. These represent the first quantitative measurement of signals from a sealed plasma panel device. A program of laboratory investigations and simulations is underway. We consider these early results to be very encouraging, especially considering that a program of device optimization has not yet been initiated.

ACKNOWLEDGMENTS

This work is sponsored by the Office of Nuclear Physics, U. S. Department of Energy and the US-Israel Binational Science Foundation.